# Correctly Rounded Functions For Vector Applications: A Performance Study

Cristina Anderson
Intel Corporation
Hillsboro, Oregon, USA
cristina.s.anderson@intel.com

Marius Cornea
Intel Corporation
Hillsboro, Oregon, USA
marius.cornea@intel.com

Andrey Stepin
Intel Corporation
Hillsboro, Oregon, USA
andrey.stepin@intel.com

Mihai Tudor Panu
Intel Corporation
Hillsboro, Oregon, USA
mihai.tudor.panu@intel.com

***Abstract*—Following recent interest in correctly rounded math library functions (as currently recommended by the IEEE 754 standard), we have designed several SIMD algorithms for one-input single precision functions and integrated them into our CPU math library; these will form the core of the first correctly rounded vector math library, to be available to users in mid-2026. To take advantage of the cross-platform bitwise reproducibility afforded by correct rounding, we adapted and evaluated a few SIMD implementations on graphics processing units (GPU). In addition, we designed and evaluated proof-of-concept SIMD implementations of two correctly rounded double precision functions.**

***Keywords—IEEE 754, correct rounding, bitwise reproducibility, vectorization.***

## I. Introduction

Correctly rounded (CR) transcendental math function implementations are relatively new. First CR implementations used an iterative approach proposed by Abraham Ziv in 1991 [1], which could be extremely slow for some inputs. His approach required multiple iterations of increasing accuracy, until an approximation of the function was found that could be proven to belong to an interval which did not contain any midpoint between two consecutive floating-point numbers – that meant that its rounding to nearest was the correct one. The need to perform an unknown number of iterations to guarantee correct rounding was eliminated by Jean-Michel Muller [2], Vincent Lefèvre [9] and others who solved the Table Maker's Dilemma problem for many double precision functions (a much more challenging problem than for smaller format functions that can be tested exhaustively). They found the arguments leading to the 'hardest-to-round' values of these functions. In turn, the hardest-to-round cases determine the required accuracy for computing an intermediate result that is guaranteed to be rounded correctly to the destination format.

There are many reasons for having and using CR math functions: the IEEE standard 754-2019 recommends many CR functions (and its revision targeted for 2029 might mandate them), they ensure bitwise reproducible results across platforms (which can improve portability), they support tighter error bounds for certain computations, and may even help with correctness proofs.

CR algorithm design has a clear accuracy goal, but other tradeoffs remain, e.g. memory footprint must be balanced against performance.

To date, many implementations have been created – all scalar, some incomplete, some work-in-progress: IBM libultim [10] which uses Ziv's method, and supports only rounding to nearest; the CNRS CR-LIBM [7] with double precision only, no longer maintained; the Sun libmcr, with only seven double precision math functions; MPFR, providing arbitrary precision arithmetic with correct rounding [11]; CORE-MATH from Inria, open-source [5]; RLIBM, open-source from Rutgers University, using a design method suitable for single precision functions [3]; the LLVM libm [12].

This paper makes an important contribution to the family of CR math libraries, by alleviating a clear drawback of CR implementations: reduced performance, particularly when compared to vectorized (and also less accurate) libraries. We demonstrate that CR functions can be implemented efficiently in vector form. We used Intel architecture SIMD (single instruction, multiple data) instructions, which generate several results simultaneously in vector registers of width up to 512 bits (up to 8 double precision, 16 single precision, or 32 float16 results). Most of our single precision implementations completely avoid branches; branches are only used for some special or very infrequent cases (such as hard-to-round cases in our double precision implementations).

The paper is organized as follows: Section II describes our general approach to algorithm design and validation for one-input single precision functions, and separately for one-input double precision functions. Section III presents our performance evaluation results in four separate sub-sections: CPU measurements for single precision functions and applications, CPU measurements for double precision (using our still experimental implementations), then GPU measurements for a few single precision functions and our double precision implementations. Section IV has a brief conclusion and plans for future work.

## II. ALGORITHM DESIGN

### A. Single Precision, One-Input Functions

Our math libraries currently include SIMD (single instruction, multiple data) implementations of the correctly rounded acosf, asinf, atanf, cosf, coshf, expf, exp2f, exp10f, expm1f, log10f, log1pf, log2f, logf, sincosf, sinf, sinhf, rsqrtf, tanf, tanhf. Our primary goal was to optimize for Intel® AVX-512 platforms, but we also included SIMD implementations for Intel® AVX2, AVX, and SSE2 platforms.

The AVX-512 implementations contain no branches, except for large argument reduction in trigonometric functions, and exact result treatment in exp10f. For other platforms, there are branches to help treat denormal inputs in log functions and rsqrt; most functions treat special inputs in the main path. Since vector gathers can be expensive, they are avoided when possible; we use small lookup tables (16 entries or less) and AVX-512 permute instructions that emulate vector gathers on small tables.

```
xd = (double)x;
// R = x – nearest_int(x*8)/8, |R|<2^-4
R = FP64_VREDUCE(xd, 0x38);
// table index in last 3 bits of d_index
d_index = FP64_ADD_RN (xd, 0x1.8p+49);
// Nd = x – R; Inf/NaN propagated to Nd
Nd = FP64_SUB_RN (xd, R);
// permute (lookup) from 8-element table
// T = (double)exp2(Nd – floor (Nd))
T = FP64_PERMUTE(&Table[0], d_index);
// poly ~ (exp2(R) -1)/R
// constant coefficients, 6 FMAs
poly = DP_FMA_RN(R, …
    …
// poly*T is exact for small |x| (T==1.0)
poly = FP64_MUL_RN(poly, T);
// result mantissa bits
result = FP64_FMA_RZ(poly, R, T);
// sticky bit computation
sticky = SHR(FP64_CMP(R, 0.0, NEQ), 63);
result = FP64_OR(result, sticky);
// scale by 2^floor(Nd); user rounding mode
// VSCALEF also handles Inf/NaN inputs
result = FP64_VSCALEF(result, Nd);
// convert to float, in user rounding mode
return (float)result;
```

Figure 1. Pseudo-code for the SIMD cr_exp2f(x) on AVX-512

Most of the functions we implemented require slightly more than double-precision accuracy level to guarantee correct rounding in all cases. The computation is performed to the required accuracy using double precision operations and round-to-nearest mode for most steps. The final steps rely on round-to-zero mode and sticky bit computation before the final conversion to single precision, in user-requested rounding mode. This approach works well on AVX-512 platforms, which have support for static rounding modes as well as other useful features for transcendental function implementation [6], and makes it easier to control the accuracy before the final rounding. Implementations ported to platforms without static rounding modes use short emulation sequences for intermediate results in round-to-zero mode.

Given that SIMD width for double precision is only half the SIMD width for single precision, future optimizations may include some fast paths based on single precision computation, with fallback to the double precision path when more accuracy is needed (similar to the approach taken in [4], [7] for scalar implementations). Platforms that lack adequate double precision support (e.g. certain GPUs) will also need a modified approach, which may include a fast path based on single precision.

Figure 1 shows a pseudo-code description of our compact SIMD cr_exp2f implementation for AVX-512 targets. For other targets, we clamp the inputs to a range that will yield normal results in double precision (but may still overflow/underflow in single precision), and emulate the few other AVX-512 specific

```
xd = (double)x;
// extract mantissa; special cases also covered
mx = FP64_GETMANT (xd, 3, 2);
// extract exponent
ex = FP64_GETEXP(xd);
// permute index for 8-entry coefficient tables
index = SHR(xd, 52 - 3);
// reduced argument:  R = 1.5*(mx-1)
R = FP64_FMS_RN(mx, 1.5, 1.5);
// exponent correction, since mx in [0.75, 1.5)
ex = ex – FP64_GETEXP(mx);
// polynomial  ~ (log2(mx) – R)/R;
// range of mx is split into 8 sub-intervals
// degree 9 polynomial in each sub-interval
c9 = FP64_PERMUTE(&c9[0], index);
c8 = FP64_PERMUTE (&c8[0], index);
poly = DP_FMA_RN(R, c9, c8);
    ….
// reconstruction: ex + R + R*poly
ex = FP64_ADD_RZ(ex, R);
result = FP64_FMA_RZ(poly, R, ex);
// no sticky bit correction is needed
// we have sufficient accuracy for log2f
// convert to float, in user rounding mode
return (float)result;
```

Figure 2. Pseudo-code for the SIMD cr_log2f(x) on AVX-512

instructions. Note that our SIMD libraries do not currently aim to set all status flags correctly (e.g. in the non-AVX512 exp2f, +Inf is clamped to a large input value that yields Overflow during the final conversion to single precision). However, the AVX-512 sequences using static rounding modes also have floating-point exceptions suppressed, so spurious exceptions will not be triggered during computation; the final conversion to single precision is allowed to set the status flags.

Figure 2 has another pseudo-code example, for the SIMD cr_log2f on AVX-512 targets. Since this sequence relies on multiple coefficient lookups from small tables (reasonably fast on AVX-512, but not on other platforms), we use one larger table and constant coefficients for the other platforms we support. In the absence of AVX-512 helper operations, the SIMD cr_log2f implementations start the computation in single precision and include an infrequently taken branch for denormal and special inputs. cr_log2f has somewhat lower accuracy requirements (below double precision level), so the non-AVX-512 sequences perform all computation in the user rounding mode.

Validation of one-input single-precision functions was done by exhaustive testing, using quad precision functions or an existing correctly rounded library for reference results.

Given that some devices (e.g. certain GPUs) have little or no double precision support, we will need to devise new algorithms to provide correct rounding (and thus bitwise reproducibility) on such platforms.

### B. Double Precision, One-Input Functions

While our primary focus so far has been SIMD single precision functions, we also wanted to explore the potential benefits of vectorization for correctly rounded double precision functions. We used cr_exp2 and cr_log for our initial experiments.

The approach we took for single precision (i.e. branch-free computation) is not feasible for double precision functions, which require much higher accuracy to guarantee correct rounding in all cases. Instead we have to use a highly accurate (but not always correctly rounded), optimized SIMD main path, and an infrequently taken secondary path that guarantees correct rounding. The secondary path is not necessarily vectorized (it was not for our experiments); it is taken when the correct rounding test applied to the main path result fails (as suggested in [4]). When this test fails, only the SIMD lanes that failed are sent to the slower path.

The main path must be accurate enough to ensure very few outputs get sent to the slow path. We considered using 64-bit integer arithmetic in place of double-double computation; this can work well for scalar computation on recent Xeon CPUs, but current SIMD support for 64-bit integer computation is limited. Our implementations use double precision computation in the fast path, but we stored the cr_log lookup table in scaled integer format in order to use one vector gather instead of two.

The fast path for cr_exp2 uses three lookup tables in double-double format (16 entries each), in order to approximate

exp2((i1*256 + i2*16 + i3)/4096.0 + R) ~ T1[i1] * T2[i2] * T3[i3] * exp2(R),

where i1, i2, i3 are integers in{0, 1, …, 15} and the reduced argument R is small enough to allow exp2(R) approximation by a short polynomial. The 16-entry tables can be read via AVX-512 fast vector gathers (permutes); however, we need 6 gathers for 3 double-double tables. The number of gathers is not ideal for platforms without permute support, but at least the small table sizes eliminate the possibility of bank conflicts when running on a GPU.

The cr_log fast path avoids multiple vector gathers by computing a short reciprocal approximation (rcp) for the input mantissa, which is first normalized to the [0.75, 1.5) range. The 128-entry lookup table stores log1p(rcp)/rcp – 1.0 in scaled integer format; the read value is converted to double-double format to continue the computation and reconstruct the log(x) approximation after also evaluating log(rcp*mantissa_x) as a polynomial. Our approach is a tradeoff between the cost of vector gathers as well as table size, and computation length.

These two implementations (using scalar correctly rounded code for the slow path) were tested with our internal test suite, as well as the sets of hard-to-round cases provided by the open-source CORE-MATH project [13].

## III. PERFORMANCE EVALUATION

### A. Single Precision Results on the CPU

For AVX-512 platforms (our main optimization target), the SIMD width is 16 for single precision and 8 for double precision operations. For correctly rounded functions, the 16 packed single precision inputs are unpacked into two sets of 8 packed double precision inputs for internal computation. This means that when compared to an equivalent scalar algorithm, one can expect at most 8X throughput speedup. In practice, SIMD algorithms use smaller tables (for faster vector gathers) and longer computation sequences, or alternatively incur the full cost of vector gathers; both scenarios limit the achievable speedup. Branch-free implementations add some (usually small) overhead for treating all inputs in one path, while the scalar sequences sometimes benefit from fast x86 arithmetic. Occasionally, the longer computation sequences may suffer from the added register pressure.

We find that on an Intel® Xeon® Platinum 8592+ processor, the geomean throughput speedup for our current set of SIMD implementations (19 functions) is 4.65X; this was measured against our own scalar implementations of the same correctly rounded functions, optimized for the AVX-512 instruction set. At the same time, these correctly rounded SIMD sequences are significantly slower than their 1-ulp accurate counterparts, which can take full advantage of the available SIMD width for single precision: the geomean for the measured slowdown is 0.42X.

Table I shows examples of our measurements (relative to SIMD correctly rounded performance for each function), along with lookup table sizes for each implementation.

We also ran a few small, synthetic benchmarks to help evaluate our new code in actual applications:

1). Spectral interpolation for a non-integer frequency lying between computed fast Fourier samples is done as a brute-force discrete Fourier transform on original time samples (as suggested in [8]). This test heavily uses sin, cos (or sincos). The AVX-512 build runs 2.78X times faster when using the SIMD cr_sincosf, versus our scalar cr_sincosf, and 2.42X faster than the 0.55-ulp scalar sincosf. The AVX2 build (run on different hardware: Intel® Core™ Ultra Processor Series 2) is 1.95X faster for the SIMD cr_sincosf versus the scalar cr_sincosf and 1.63X faster than the 0.55-ulp scalar sincosf. Using separate cr_sinf and cr_cosf calls, the vectorized AVX-512 build is 2.76X faster than our scalar CR functions (and 4.06X faster than CORE-MATH); the AVX2 build is 1.95X faster than our scalar CR functions (and 2.39X faster than CORE-MATH).

TABLE I. Examples of AVX-512 single precision functions and their performance relative to the SIMD CR version (lower ratio means faster)

| CPU Function | SIMD width | Reciprocal Throughput Cycle Ratio | Lookup Table Size |
|---|---|---|---|
| **SIMD atanf, 1 ulp** | 16 | 0.36 | 0.25 KB |
| **SIMD cr_atanf** | 16 | 1 | 0.125 KB |
| **scalar cr_atanf** | 1 | 4.2 | 2 KB |
| **SIMD sinhf, 1 ulp** | 16 | 0.48 | 0.5 KB |
| **SIMD cr_sinhf** | 16 | 1 | 0.125 KB |
| **Scalar cr_sinhf** | 1 | 4.05 | 1 KB |

2). Quick computation of 2-ulp complex expf, complex logf, complex sinf (valid inputs only) based on our available single precision correctly rounded functions; measured for AVX-512, with real and imaginary input fields in [-40,40]:

- cexpf (using cr_expf, cr_sinf, cr_cosf) is 5X faster when vectorized (and 7.67X faster, compared to CORE-MATH)

- clogf (using cr_logf, cr_log1pf, cr_atanf as well as division and square root instructions) is 3.21X faster when vectorized (and 5.27X faster, compared to CORE-MATH)

- csinf (using cr_coshf, cr_sinhf, cr_cosf, cr_sinf) is 4.89X faster when vectorized (and 6.7X faster, compared to CORE-MATH)

In some cases, benchmark measurements (relative to scalar performance) outperform expectations based on individual function timing. This is because our SIMD sequences are branch-free, while some scalar implementations have a few different computation paths and can suffer from branch penalties that are not always captured in measurements using a uniform distribution over a limited input range. (The scalar implementations in our Intel LIBM generally have few branches, mostly limited to special or fast exit cases. We find that while our LIBM CR implementations are comparable in performance to the open-source CORE-MATH on individual function tests, CORE-MATH is more prone to branch penalty slowdown when used in a larger benchmark, such as the applications described above.)

### *B. Double Precision Results on the CPU*

We evaluated our AVX-512 SIMD cr_log against our own scalar cr_log. Since we do not currently have a scalar implementation of cr_exp2, our experiment used the open-source CORE-MATH cr_exp2 [5] as the reference for performance evaluation, as well as for the infrequently taken accurate path (call to an accurate scalar function, i.e. a "callout" in the following text).

The relative cost of callouts to the very accurate path is higher for SIMD implementations, since an entire packed vector computation is disrupted when one element requires the accurate path. Our SIMD cr_exp2 has 20% lower throughput when the accurate path is plugged in, compared to running the main path only. The SIMD cr_log can afford less frequent callouts to the accurate path when the inputs are sufficiently far from 1.0; it still loses ~21% of main path throughput with uniformly distributed inputs over [0.5, 2], but less than 10% for significantly wider ranges. We expect the numbers can be somewhat improved by ensuring the callouts are built for each target architecture; infrequent callouts are currently meant to be shared by all CPU targets, so they are not optimal for AVX-512 platforms.

Table II shows our performance results for exp2 and log, as ratios relative to each correctly rounded SIMD implementation. We used uniformly distributed inputs in [-20,20] for exp2, and [0.125, 8.0] for log.

TABLE II. AVX-512 double precision exp2, log and their performance relative to the SIMD CR version (lower ratio means faster)

| CPU Function | SIMD width | Reciprocal Throughput Cycle Ratio | Lookup Table Size |
|---|---|---|---|
| **SIMD exp2 1 ulp** | 8 | 0.56 | 0.25 KB |
| **SIMD cr_exp2** | 8 | 1 | 0.75 KB |
| **CORE-MATH cr_exp2** | 1 | 5.61 | 2 KB |
| **SIMD log, 1 ulp** | 8 | 0.55 | 0.125 KB |
| **SIMD cr_log** | 8 | 1 | 1 KB |
| **Scalar cr_log (Intel)** | 1 | 4.55 | 2 KB |

### *C. Single Precision Results on the GPU*

We adapted our SIMD cr_expf, cr_logf, cr_sincosf to work on GPUs. The GPUs we selected for this experiment do not support static rounding modes, and dynamic rounding mode changes from high-level code are unwieldy. As such, we ran the experiment with computation and final rounding in the default rounding mode only (round-to-nearest even).

The first GPU we selected (Intel X$^{e}$-HPC, or "GPU 1") has very good double precision support; basic double precision operations have the same throughput as their single precision counterparts. As expected, our correctly rounded functions are only a little slower than the less accurate, 1-ulp versions available in our GPU libraries.

The second GPU (Intel Arc B580, or "GPU 2") supports double precision at much lower throughput than single precision. As a result, our correctly rounded functions (relying almost exclusively on double precision computation) are much slower than their 1-ulp counterparts, which use single precision computation. Platform-specific optimization will be needed to boost performance of single precision CR functions on GPU 2.

Table III shows the relative performance of correctly rounded functions versus 1-ulp functions on these two GPUs.

### D. *Double Precision Results on the GPU*

The SIMD cr_exp2 and cr_log implementations we tested on AVX-512 were adapted for the GPU (in round-to-nearest mode) and compared against their 1-ulp counterparts from GPU libraries.

On GPU 1, the SIMD main paths are similar in performance to the 1-ulp sequences. On GPU 2, the lookup-table implementation used for the cr_log main path is 1.4X times faster than the polynomial-based 1-ulp log (leading to a faster overall cr_log); the cr_exp2 main path (containing multiple small gathers) is slower than the 1-ulp exp2.

Table IV summarizes our results for these GPU functions, based on throughput measurements using uniformly distributed inputs in [-20,20] for exp2, and [0.125, 8.0] for log.

TABLE IV. Relative performance of correctly rounded versus 1-ulp implementations; lower ratio means correctly rounded function is slower

| GPU Function | GPU 1 | GPU 2 |
|---|---|---|
| **exp2** | 0.73 | 0.58 |
| **log** | 0.71 | 1.16 |

## IV. DISCUSSION AND FUTURE WORK

Our experiments on one-input single and double precision functions show that correct rounding can be supported with reasonable performance on multiple platforms. While there can still be significant performance loss when compared to vectorized 1-ulp implementations, the tradeoff will be acceptable for users that require cross-platform bitwise reproducibility, or the best possible elementary function accuracy for a given floating-point format.

Since some modern GPUs de-emphasize double precision support and some smaller devices only support single precision, we will look into alternative implementations for correctly rounded single precision functions on such platforms. For example, we expect that performance of single precision CR functions on the Intel Arc B580 (“GPU 2”) will be improved by employing 32-bit operations in internal computation.

TABLE III. Relative performance of correctly rounded versus 1-ulp implementations; lower ratio means correctly rounded function is slower

| GPU Function | GPU 1 | GPU 2 |
|---|---|---|
| **expf** | 0.93 | 0.24 |
| **logf** | 0.76 | 0.19 |
| **sincosf** | 0.81 | 0.20 |